\documentclass[journal,12pt,draftclsnofoot,onecolumn]{IEEEtran}

\ifCLASSINFOpdf
 
\else
  
\fi

\usepackage{multicol}
\usepackage{lipsum}
\usepackage{setspace}
\usepackage{amsmath}
\usepackage{cases}
\usepackage{cite}
\usepackage{amsfonts}
\usepackage{amssymb}
\usepackage[linesnumbered,ruled,vlined]{algorithm2e}

\usepackage{array}
\usepackage[margin=1.0in]{geometry}
\usepackage{amsthm}
\usepackage{empheq}
\renewcommand*{\stackrel}{%
\mathrel\bgroup\stack@relbin
}

\ifCLASSINFOpdf
   \usepackage[pdftex]{graphicx}

  \graphicspath{{../pdf/}{../jpeg/}{../eps/}}
  \DeclareGraphicsExtensions{.eps,.pdf,.jpeg,.png}
\else
   \usepackage[dvips]{graphicx}
   \graphicspath{{../eps/}}
   \DeclareGraphicsExtensions{.eps}
\fi
\usepackage{epsfig}
\usepackage{fmtcount}
\usepackage{mwe}    
\usepackage{subfig}

\usepackage{array}
\usepackage{url}
\usepackage[linesnumbered,ruled,vlined]{algorithm2e}
\usepackage{multicol}
\usepackage{float}
\usepackage{lipsum}
\usepackage{color}

\usepackage{mathtools}  
\usepackage{tabulary}
\usepackage{booktabs}
\usepackage{stfloats}

\begin{document}
\title{{\fontsize{17}{17}\selectfont On Metropolis Growth}}

\author{Syed~Amaar~Ahmad, \emph{Member, IEEE}
\thanks{Syed Amaar Ahmad received his PhD in Electrical Engineering from Virginia Tech. Currently he is also working on crash avoidance through vehicular communications in the Metro-Detroit area. Email: syed.a.ahmad@ieee.org.}}

\markboth{}
{Shell \MakeLowercase{\textit{et al.}}: Bare Demo of IEEEtran.cls for Journals}

\maketitle

\thispagestyle{plain}
\pagestyle{plain}

\begin{abstract}
We consider the scaling laws, second-order statistics and entropy of the consumed energy of metropolis cities which are hybrid complex systems comprising social networks, engineering systems, agricultural output, economic activity and energy components. We abstract a city in terms of two fundamental variables; $s$ \emph{resource cells} (of unit area) that represent energy-consuming geographic or spatial zones (e.g. land, housing or infrastructure etc.) and a population comprising $n$ mobile units that can migrate between these cells. We show that with a constant metropolis area (fixed $s$), the variance and entropy of consumed energy initially increase with $n$, reach a maximum and then eventually diminish to zero as a \emph{saturation} threshold is reached. These metrics are indicators of the spatial  mobility of the population. Under certain situations, the variance is bounded as a quadratic function of the mean consumed energy of the metropolis. However, when population and metropolis area are endogenous, growth in the latter is arrested when $n\leq\frac{s}{2}\log(s)$ due to diminished population density. Conversely, the population growth reaches equilibrium when $n\geq {s}\log{n}$ or equivalently when the aggregate of both over-populated and under-populated areas is large. Moreover, we also draw the relationship between our approach and multi-scalar information, when economic dependency between a metropolis's sub-regions is based on the entropy of consumed energy. Finally, if the city's economic size (domestic product etc.) is proportional to the consumed energy, then for a constant population density, we show that the economy scales linearly with the surface area (or $s$). 
\end{abstract}


\IEEEpeerreviewmaketitle

\section{Background}
\IEEEPARstart{A}{n} urban metropolis represents a large city or population concentration that is a significant economic,commercial, political, and cultural center within a country or region, and has vital regional or international links. A metropolis therefore represents a sprawling and unpredictable natural structure that is a complex system. Capturing the features of such systems is often difficult, and appropriate mathematical models are needed to understand their working. Consisting of social networks, engineering structures, energy generators and urban agriculture, a metropolis is therefore an ultimate form of a complex system.

The growth of metropolis cities in terms of population, occupied area and economic activity are of critical interest to urban planners, governments and businesses. The population growth of cities is economically important in itself helping to drive the national growth by virtue of sheer magnitude. Investments in constructions, housing, infrastructure and energy sources helps accommodate the rising population of cities. The economic and urban development literature on the connection between the growth of cities, population, spatial expanse (i.e. surface area) and per capital income has been crisply summarized in \cite{Met10}. This work reviews key theories for urban growth that build atop the seminal work \emph{The Isolated State} by 19th century German economist von Th{\"u}nen \cite{thunen}.

Urban development researchers have long known that transportation costs are a basic factor of the population of cities and underlying patterns of land usage \cite{Met10}. A \emph{linear monocentric} urban city is based on a model where commuting costs increase linearly with distance to the city center for workers. The literature for economic analysis on urban centers uses this monocentric model for housing and transportation. The results it draws also assumes that the workers in the population are homogeneous and perfectly mobile. As \cite{Met10} argues, the studies on the growth of cities so far has not been able to address how growth and technological progress mutually affect each other.

In this paper, we study a metropolis as a \emph{complex system} \cite{Met4} from a new perspective, where large-scale system behavior emerges from the underlying behavior of individual small-scale components. In complex systems, individual small components have behaviors that can aggregate to a more complex model. A metropolis exhibits a complexity where individual members of a large population move across its region under arguably some simple rules. 

A common theme for complex systems is how do they \emph{scale}. The notion of scaling may have various connotations; ranging from wireless systems \cite{Met7} to micro-organisms, and from gas thermodynamics to human civilizations \cite{Met6}. Another aspect of complexity pertains to self-similarity that is a recurrent theme in biological systems \cite{plantShoot}. In this paper, the term \emph{scale} is used to describe three situations; (1) the growth in the population size given a constant metropolis spatial region, (2) increase in the number of underlying spatial region given a fixed population size, and (3) increase in the number of physically isolated (or otherwise) subsets of individual units given an overall constant population size and spatial region.  

Existing work on the energy consumption on urban population centers uses empirical data for a sample set of US Zip codes (see \cite{zipEnergy} and its references). The study in \cite{zipEnergy} investigates the affect of socioeconomic and demographic patterns on the residential energy utilization patterns.  It uses regression to show that total population and energy use are strongly correlated and that areas with low population densities tend to consume significantly more energy per capita than those with high population densities. The empirical relationship between the economic size and energy consumption have also been explored in \cite{Karanfil_ecoElec,rand_enisil}. 

The energy consumption of a metropolis' sub-regions can also help identify the internal dynamics and economy of the city. A key question is how representative is the energy consumption of a sub-region for the remaining portion of the city.
Self-similarity \cite{Met4} sheds light on the fact that complex systems exhibit patterns where a small component may provide representation for the larger self. Determining the energy consumption characteristics of a metropolis may also be useful in determining the economic dependencies or self-similarity of sub-regions of the city. 

A complex system's internal function is exhibited by the information between its sub-components, each operating with a set of simple rules. Developments in information theory \cite{coverthomas} provide a powerful tool for quantifying the complexities and internal dependencies of system. The \emph{mutual information} reflects the dependencies between various sub-components of a system. Each sub-component provides inferences on the system's dynamics based on statistical or probabilistic relationships.

The work in \cite{Met7} explores the concepts of \emph{multi-scalar information}, in terms of how a system's joint information is spread over each of its irreducible smaller components. \cite{Met7} shows that there is a marginal utility of information depending on the scale. It is also concerned with  whether the system's components (e.g. intelligent machines, particles or social groups etc.) are largely independent or strongly interdependent.

\subsection{Contributions}
We treat the metropolis as a model with only two variables of population and energy-consuming \emph{resource cells} (or simply cells) that could be a set of housing units, buildings, city blocks or arbitrarily large areas. A cell also represents some arbitrary geographical area, land resources or composite residential and industrial infrastructure. The consumed energy of each cell is presumed to be a hybrid mix of solar, fossil fuel or alternative sources. In this paper we present the following new concepts:\\

1. We propose a topological model of the population distribution in a single metropolis city. The population units are mobile and move back and forth between the cells. We treat the population distribution of the city as a random field, with an \emph{Euler characteristic}-like function, we can represent the overall energy consumption of the city fluctuating with the population distribution. The randomness stems from the underlying actual movement of people and goods throughout the city. As the population moves about the metropolis, the cells with the densest population change. In this model, the notion of which cells represent the suburbs or the city center, is replaced by how many cells have the highest concentration. Therefore, while the specific cells with the highest population may randomly change, the number of cells with a high population may be invariant. 

Random field theory is often used for brain imaging \cite{randFieldTheory} to identify regions of high activity. Inspired from this, the key parameter in our model is based on the number of cells with high population concentrations consuming large energy. The advantage of adopting this approach for a metropolis is that it reflects  the dynamic traffic and energy consumption patterns without being tied to specific mobility models. Under our approach, the population randomly move across any cell of metropolis. This allows us to determine the energy statistics when the population has maximal mobility and contrast against when mobility patterns are constrained to smaller areas. Our approach thus sheds light on arbitrary population mobility models. \\

2. We derive scaling laws on the endogenous growth of population and the metropolis area (as counted by the number of cells) in terms of two complementary inequalities. In the first law, namely the \emph{cell growth}, the population is fixed but the number of cells (or area) increase until the metropolis population density drops below a certain threshold. We show that this threshold corresponds to when the exterior area adjacent to the city's boundary equals the average unoccupied area in its interior. In the second approach, namely the \emph{population growth}, the number of cells is constant but the population increases only if a certain minimum fraction of the metropolis area is neither under-populated nor over-populated. We define two simple inequalities that denote the Pareto-optimal equilibria which determine whether only population or only metropolis area can increase.\\

3. Given a constant metropolis area, we evaluate the mean and the variance of the consumed energy statistics as a function of population. As the population increases for a constant metropolis area, the variance and entropy in consumed energy eventually diminishes to zero as the corresponding mean consumed energy approaches an asymptotic limit. With initial increases in the population, there is increasing variation in consumed energy. However, after a certain point the population concentration for most cells becomes too high leading to saturation in consumed energy. We show that the variance of consumed energy is bounded by a quadratic function of the city's mean consumed energy under certain circumstances. If the consumed energy is representative of a city's economic size, then we also show that given a constant population density, as the city's area grows, its domestic product scales linearly with this area.\\

4. We show that the variance (and entropy) of the consumed energy decline when the population is proportionally divided into smaller, discontiguous and isolated sub-regions than as one whole. This implies that the variance or entropy represent the degree of mobility (spatial freedom) of the population.\\

5. From an information theoretic perspective, the mutual information of the consumed energy between any given sub-region and the remaining area of the metropolis diminishes if the area of the two zones is disproportional. If consumed energy is a measure of the economy, it means that smaller sub-regions are a less representative sample of the economic characteristics of the whole city. Our results reinforce the results in \cite{Met7} that the marginal utility of information declines with sub-scaling. \\

The structure of this paper is as follows. In Section II, we present a system model to represent a metropolis. In Section III we then postulate scaling laws and equilibrium points for the growth of population and metropolis spatial spread. The second order statistics (i.e. mean and variance) of the consumed energy are analyzed in Section IV. We then explore the entropy properties of the consumed energy and present results on the mutual information between sub-scales of the metropolis in Section V. Finally, Section VI provides conclusions and directions for future work. 

\section{System Model}
Let a metropolis be partitioned into $s$ identical geographical cells, each of unit area, representing resource cells such as land, housing or electric grid infrastructure. The partitioning represents an equal division of the metropolis area into cells where each may consume the same maximum energy. The cells are located in a pattern which may be square, circular or some other regular shape. The region is occupied by $n$ \emph{population units} where each unit may be an individual person or a large grouping of persons. Without loss of generality, we assume that each of $n$ population units can randomly move between any of the $s$ cells with uniform distribution in successive units of time (or iterations). A cell may be occupied by zero, one or more population units in any iteration. A uniform distribution for mobility represents the case when the population units have maximal mobility and may migrate between all cells freely. We assume \emph{ergodicity}, where the iterated ensemble statistics denote the \emph{short-term} temporal population movements.

\begin{figure}[t]
     \subfloat[Iteration 1\label{subfig-1:dummy}]{%
       \includegraphics[width=0.45\textwidth]{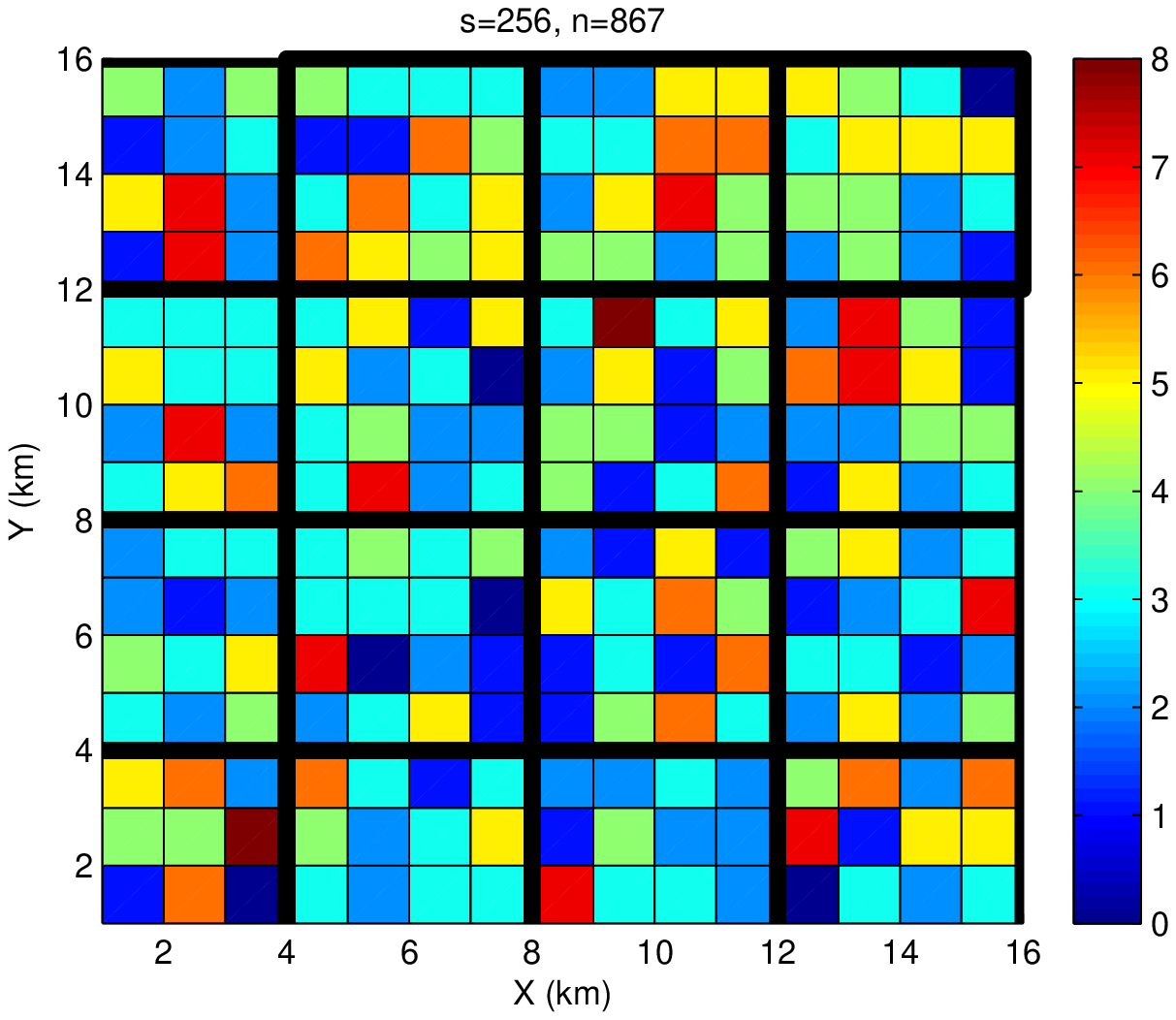}
     }
     \hfill
     \subfloat[Iteration 2\label{subfig-2:dummy}]{%
       \includegraphics[width=0.45\textwidth]{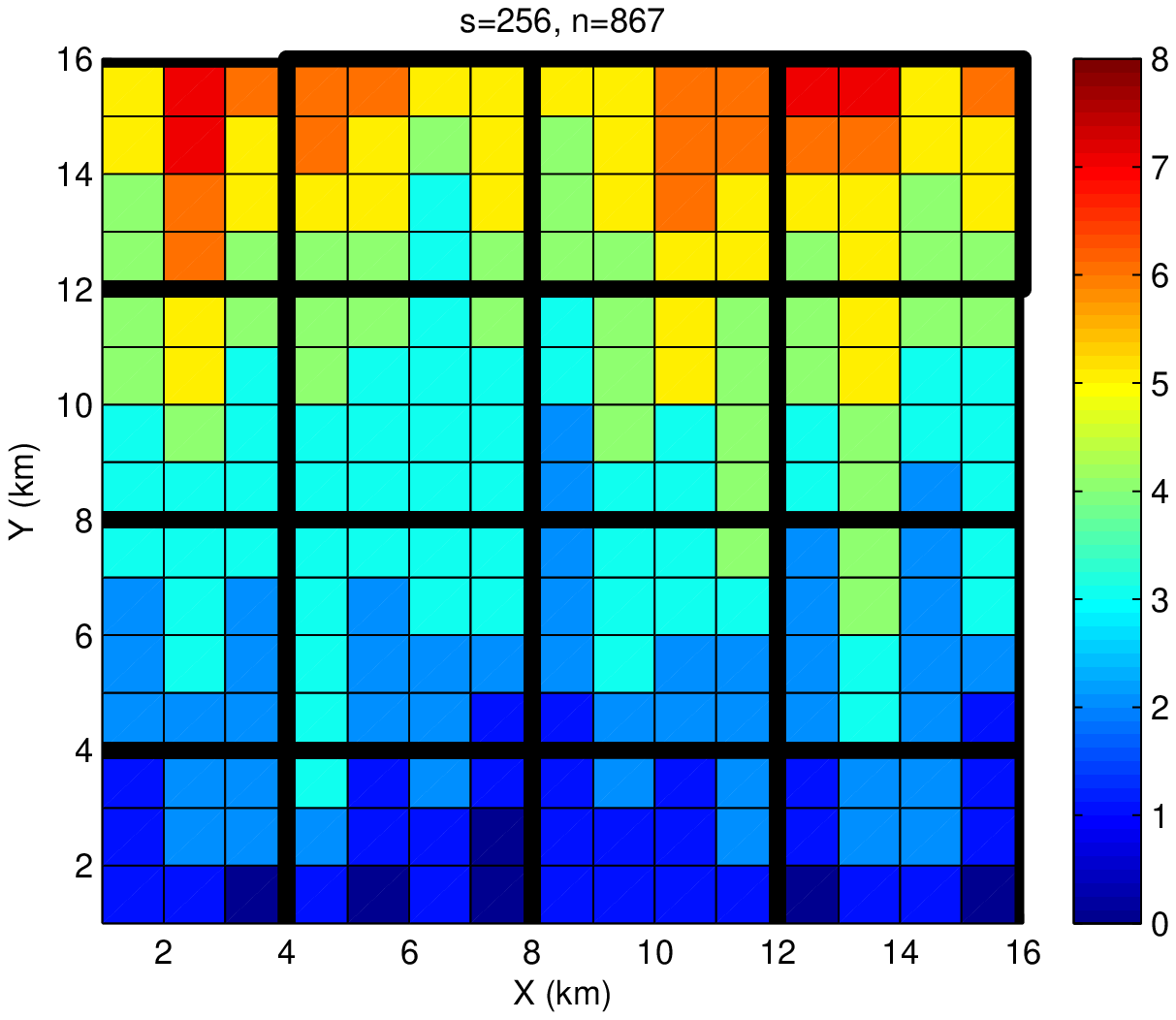}
     }
     \caption{In an epoch, the population randomly moves between cells in the contiguous 2-D metropolis region with $k=16$ sub-regions. A deep blue-colored cell has no population unit.}
     \label{fig:furt1} 
   \end{figure}

We let $L_j: L_j\leq s$ be a random variable that denotes the number of cells occupied by $j$ population units. For instance, $L_0=80$, $L_1=100$ and $L_2=50$ indicates that $80$ cells have no population units, $100$ cells are each occupied by a single population unit, $50$ cells are each occupied by two population units and so on. To reiterate, a population unit may comprise an arbitrarily large set of individuals. Hence, $L_0$ essentially represents the number of {quasi-empty} or unoccupied cells with little or no population. In Fig. \ref{fig:furt1}, for example, the total number of deep blue colored cells equals $L_0$.

When both $n$ and $s$ are large, a Poisson random distribution with intensity parameter (i.e. metropolis population density) 
\begin{equation}
\lambda = \frac{n}{s},
\end{equation}  
is a good approach to determine the average number population units occupying the cells (as validated by Monte-Carlo simulation). The average number of cells occupied by $j$ population units is denoted as   
\begin{equation} 
\label{eq2}
\begin{split}
{\bf{E}}\left[L_j\right] &= s\frac{\lambda^{j}}{j!}\exp\left(-\lambda\right)\\
&=s\frac{(n/s)^{j}}{j!}\exp\left(-\frac{n}{s}\right),
\end{split}
\end{equation}
where ${\bf{E}}[\bullet]$ is the expectation operator. To reiterate, while the population units randomly move back and forth between the $s$ cells, ${\bf{E}}\left[L_j\right]$ denotes the average number of cells that will each be occupied by $j$ population units.

\begin{figure}[t]
     \subfloat[\label{subfig-1:dummy}]{%
       \includegraphics[width=0.45\textwidth]{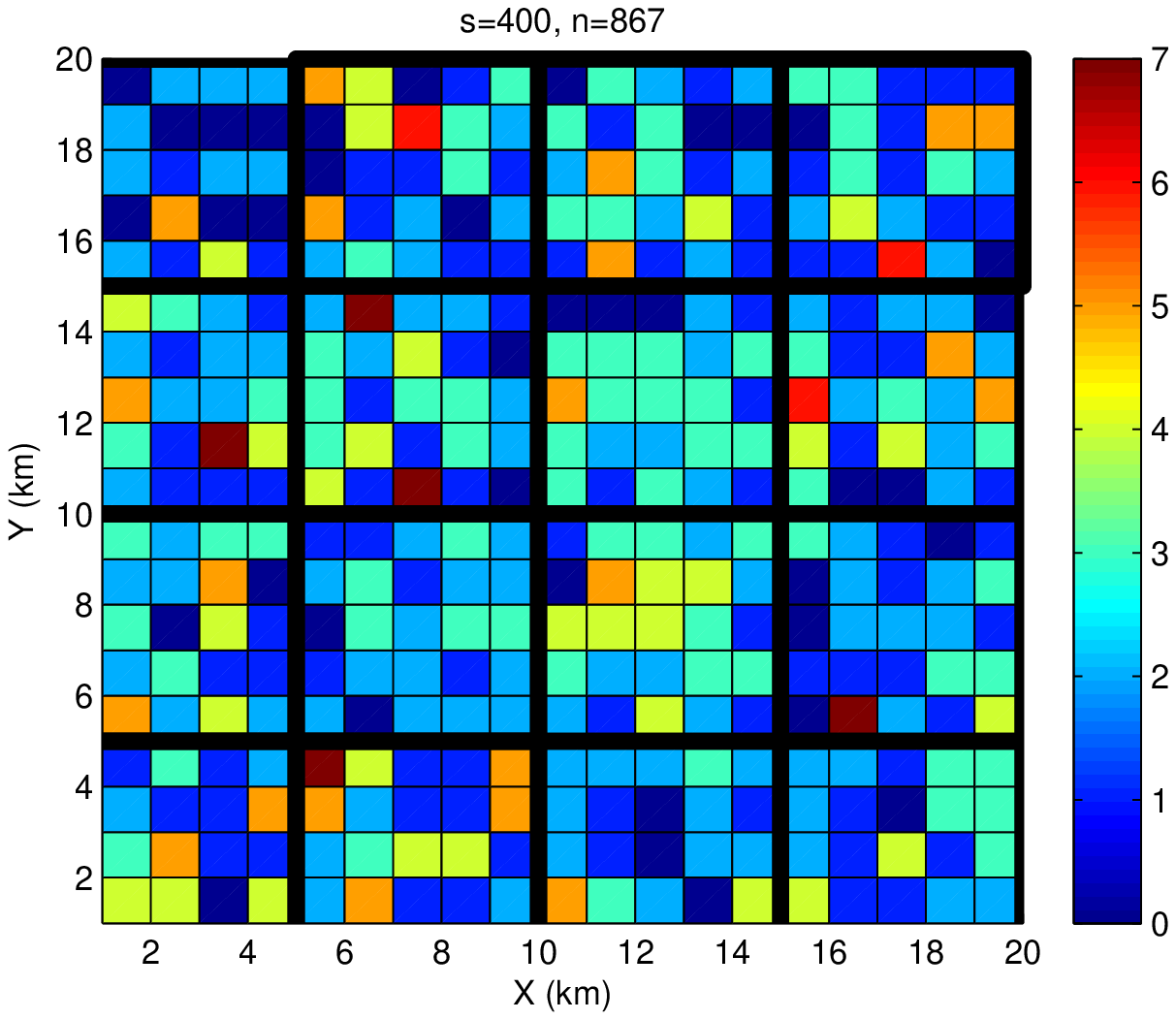}
     }
     \hfill
     \subfloat[\label{subfig-2:dummy}]{%
       \includegraphics[width=0.45\textwidth]{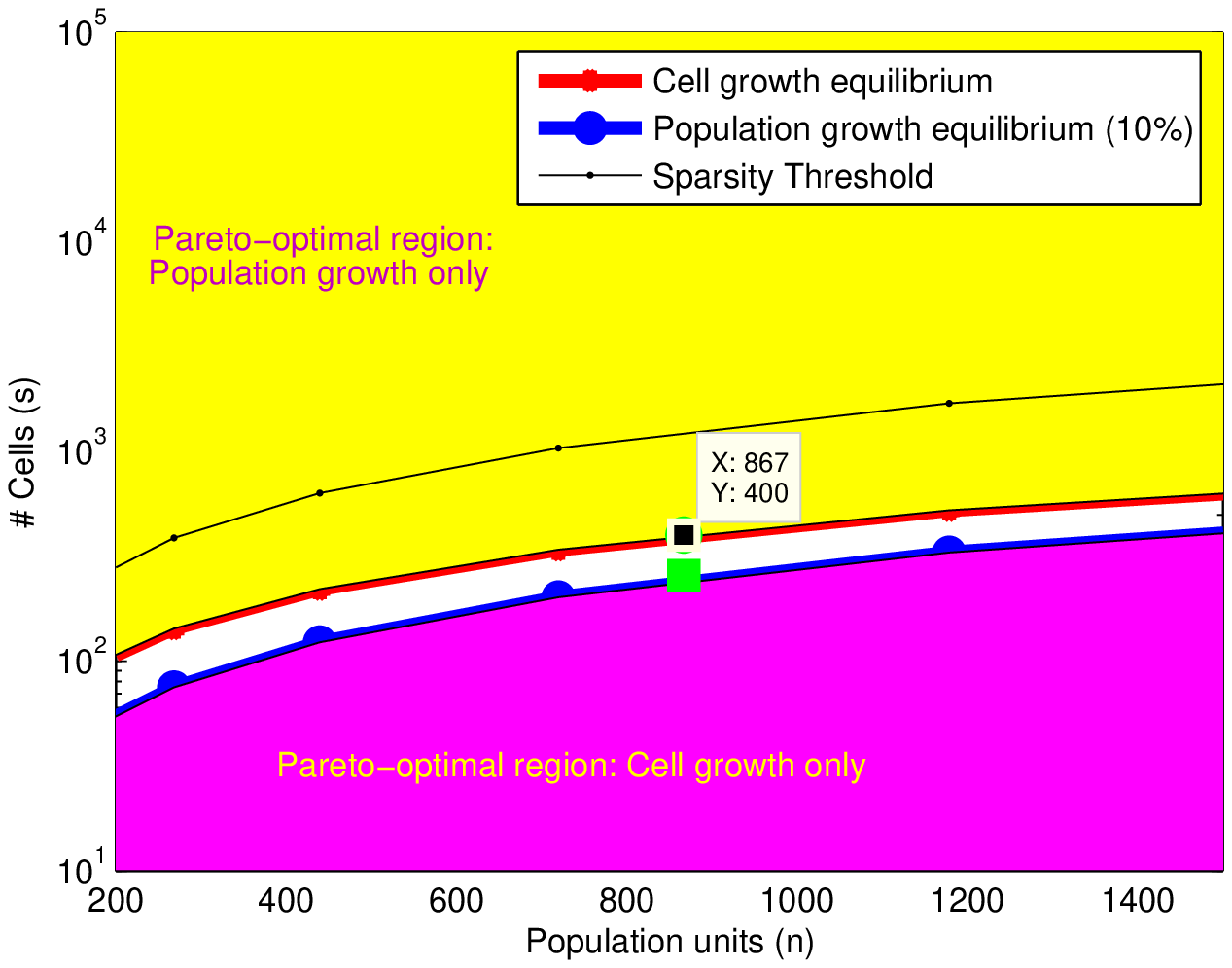}
     }
     \caption{(a) The metropolis area grows from 256 to 400 cells, where the area adjacent to its perimeter equals the average unoccupied area (deep blue cells) within. (b) The equilibria relationship between population and cells is depicted here.}
     \label{fig:furt2}
   \end{figure}

\subsubsection{Consumed Energy}
\begin{figure}[t]
     \subfloat[\label{subfig-1:dummy}]{%
       \includegraphics[width=0.45\textwidth]{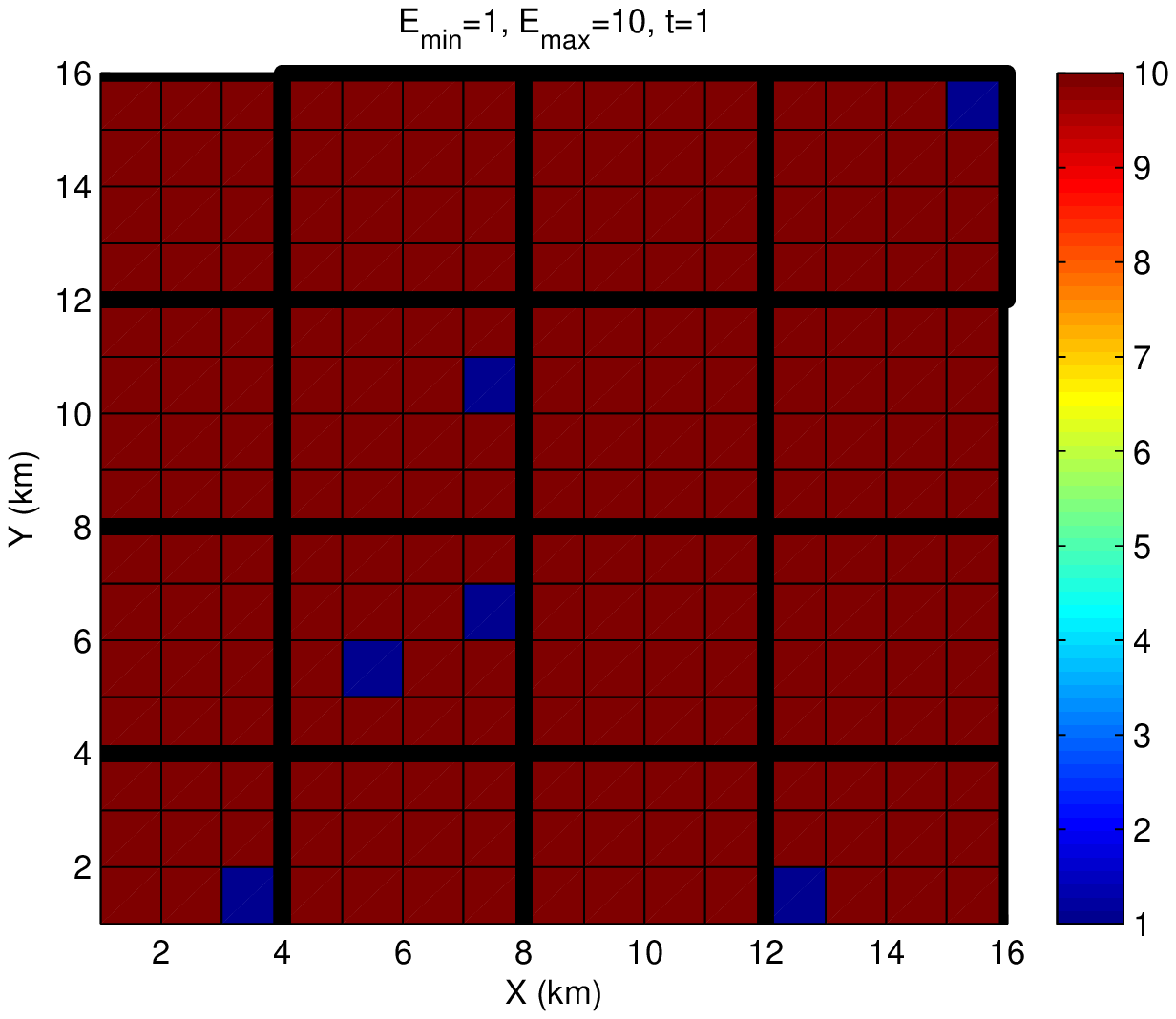}
     }
     \hfill
     \subfloat[\label{subfig-2:dummy}]{%
       \includegraphics[width=0.45\textwidth]{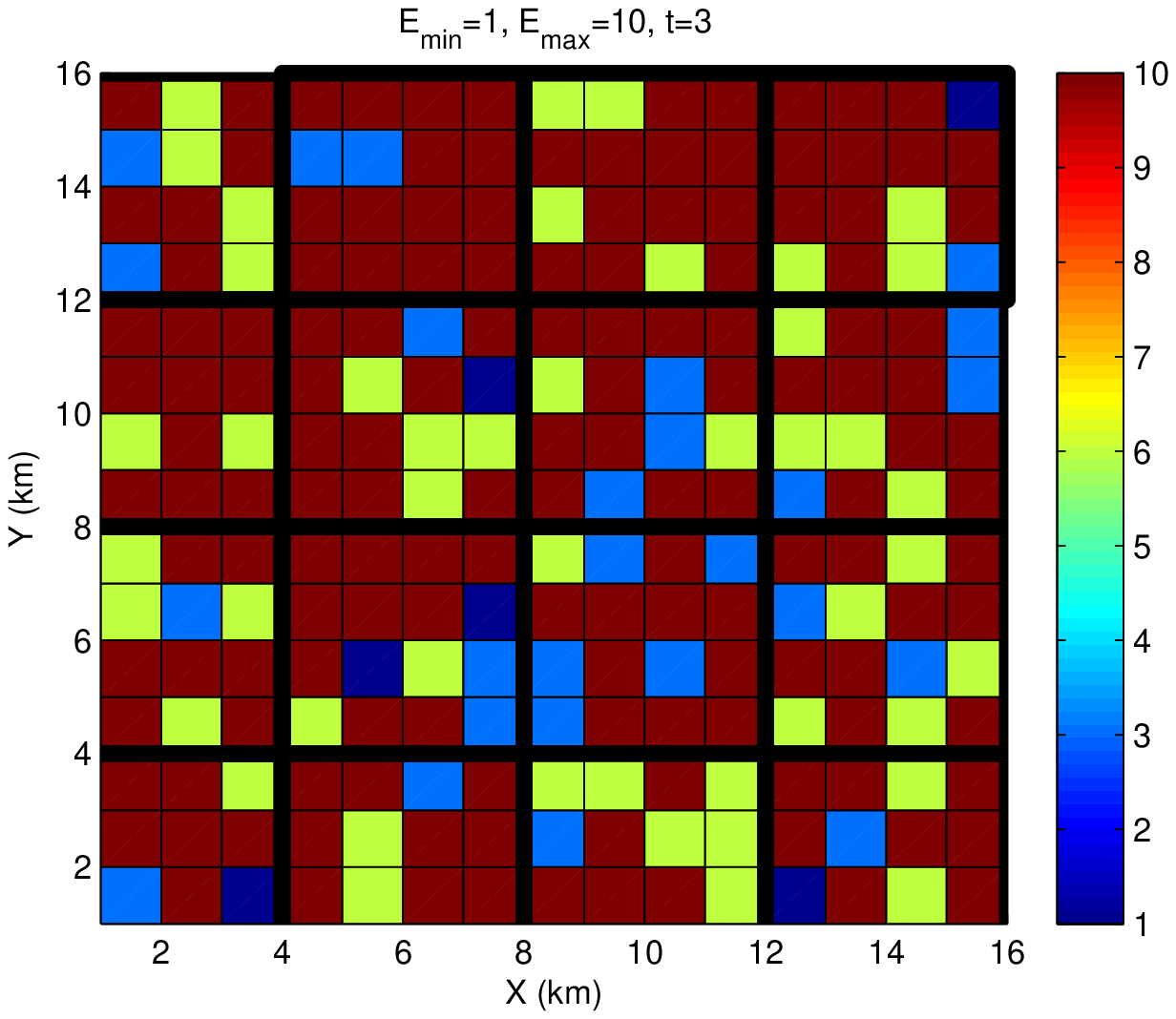}
     }
     \caption{Given (a) 2 and (b) 4 possible quantized energy levels per cell, the equivalent heat map of the energy consumption of the metropolis for the population distribution in Fig. \ref{fig:furt1} (a).}
     \label{fig:energyheatmap}
   \end{figure}

The energy consumed by each of the $s$ cell comprises an arbitrary mix of solar, fossil fuel or alternative sources. This hybrid mix is presumed identical for each cell, and is calibrated to scale between a minimum $E_{\min}$ and a maximum of $E_{\max}$ units of energy. The lower limit is due to an irreducible energy consumption per cell, whereas, the upper limit is due to a constraint on how much energy a cell may maximally avail. For instance, if a cell represent a housing unit, then even with full occupancy, the consumed energy of such a unit will not exceed some maximum supplied energy. Likewise, the units will always require some minimum energy for maintenance or sustenance. Between these two limits, however, the consumed energy depends on the number of occupying population units. In an iteration, a simple energy consumption model for each cell $i \in \{1,2,\cdot, s\}$ as a function of the number of population units $j$ occupying it, is as follows 
\begin{equation} 
\label{eq1}
M_i =
\begin{cases}
\frac{E_{\max}-E_{\min}}{t}j + E_{\min} \mbox{, if $j<t$}\\
E_{\max}\mbox{, if $j\geq t$},
\end{cases}
\end{equation}
where $t$ is a saturation threshold. The consumed energy $\mathcal{E}$ of the metropolis in any given iteration is 
\begin{equation} 
\label{eq3}
\mathcal{E}=\sum_{i=1}^s M_i =
\left(\sum^{t-1}_{j=0}L_j\left(\frac{E_{\max}-E_{\min}}{t}j + E_{\min}\right)+\sum^{s}_{j=t}L_jE_{\max}\right).
\end{equation}
As the population moves about, the random variables $\mathcal{E}$ changes accordingly. A depiction of the energy consumption variations is shown in Fig. \ref{fig:energyheatmap} for the population distribution in Fig. \ref{fig:furt1}(a). A higher saturation $t$ threshold leads to greater variation in the energy consumption distribution.

\subsubsection{Contiguous Region or Discontiguous (Isolated) Sub-regions}
The metropolis can be viewed as comprising $k: 1 \leq k \leq s$ sub-regions, where each sub-region comprises a fixed set of $\frac{s}{k}$ cells. For tractability, we assume that $k$ is such that $\frac{s}{k}$ is a natural number. If these sub-regions are part of a contiguous region,  all the $n$ population units can move between all the $s$ cells. Conversely, if all these $k$ sub-regions are discontiguous, then the population in each is essentially isolated from the others.  In this latter case, we presume that each sub-region then has $\frac{n}{k}$ population units also, where the randomness in population mobility is independent and identically distributed (iid) with uniform distribution within each sub-region of $\frac{s}{k}$ cells. 

\section{Scaling Laws}
In our proposed model, the scaling in both the surface area and the population of the metropolis occur over epochs of much longer time duration than the short-term iterations. We let the metropolis be characterized by its area and population in a  given \emph{epoch} by $(n,s)$, where the population units follow the short-term iterative behavior outlined earlier. In the next epoch, the pair $(n,s)$ representing the metropolis may evolve to different values.

We denote the average number of unoccupied (or quasi-empty) cells as $A_e= {\bf{E}}\left[L_0\right]$ and the average number of occupied cells as $A_o = s - A_e$, respectively. Using \eqref{eq2}, these variables equal 
\begin{equation} 
\label{eq2x}
A_e=s\frac{(n/s)^{0}}{0!}\exp(-\frac{n}{s}) = s\exp\left(-\frac{n}{s}\right)\\
\end{equation}
\begin{equation} 
\label{eq2xx}
A_o = s-A_e = s\left(1-\exp\left(-\frac{n}{s}\right)\right)
\end{equation}

\subsection{Sparsity Threshold}
Consider the situation when the number of population units is such that the average number of occupied cells equals the average number of unoccupied cells. Equivalently, we have
\begin{equation} 
\label{eq8x}
\begin{split}
A_e&= A_o\\
n &= s\log(2)
\end{split}
\end{equation}
The relationship in \eqref{eq8x} can be generalized to when the average number of unoccupied cells is more than $\frac{s}{k}$ cells,  
\begin{equation} 
\label{eq8xx}
\begin{split}
A_e&\geq \frac{k}{s}\\
n&\leq s\log(k),
\end{split}
\end{equation}
where $1 \leq k \leq s$. Given the pair $(n,s)$, if the inequality in \eqref{eq8xx} is satisfied, then the metropolis is deemed sparsely populated. Specifically, \eqref{eq8x} represents the sparsity threshold of the metropolis.

\subsection{Cell Growth Equilibrium}
We consider the situation when the population is fixed but the metropolis area, as denoted by the number of cells $s$, can increase in successive epochs. Under this approach, in each epoch, there is expansion of the metropolis area provided that the population is sufficiently large.  

The increase in the number of cells occurs along the edge of the metropolis area and the scaling of the area is therefore proportional to the metropolis' perimeter. Given a constant population $n$, we postulate that the expansion stops when the area gained through new cells along the perimeter equals the average unoccupied area within the metropolis. Throughout the expansion, all new cells remain contiguous to the city. 

Let the scaling factor in the number of cells be $k:k>1$, where $s$ increase to $ks$. The unoccupied metropolis area must not become more than the area of the cells along the perimeter
\begin{equation} 
\begin{split}\label{eq9a}
A_e&\leq ks-s\\
sk\exp\left(-\frac{n}{sk}\right)&\leq  s\left(1-\frac{1}{k}\right)\\
n&\geq sk\log(1-\frac{1}{k}).
\end{split}
\end{equation} 
Thus, there is spatial growth of the metropolis area by a factor of $(1-\frac{1}{k})$. For a square region, such as shown in Fig. \ref{fig:furt1}, each of the $s$ cells is unit area ($1 \times 1$). For example, a city in the square grid will expand from $16 \times 16$ in successive epochs to $17 \times 17$, $18 \times 18$, $19 \times 19$ and finally to a $20 \times 20$ square region as shown in Fig. \ref{fig:furt2}a.

Hence, the increase in the number of cells at the perimeter in one scaling increment along each axis is about $4\left(\sqrt{s}+1\right)$. Specifically, here $k=1+\frac{4}{\sqrt{s}} + \frac{4}{s}$, and the relationship in \eqref{eq9a} is then
\begin{equation} 
\label{eq9x}
\begin{split}
n&\geq sk\log(1-1/k)\\
n&\geq s(1+\frac{4}{\sqrt{s}}+\frac{4}{s})\left(\log\left(4+\sqrt{s}\right)-\log(4)\right)
\end{split}
\end{equation}
The shape of the metropolis region can affect the expression in \eqref{eq9x}. For instance, with a circular
metropolis region, the increase in the number of cells in one scaling increment is each direction by one along the circumference is approximately $2\sqrt{\pi s}$ with $k \approx 1 + 2\sqrt{\frac{\pi}{s}}$. This modifies the expression to
$n\geq s\left(1 + 2\sqrt{\frac{\pi}{s}}\right)\left(\log\left(2\sqrt{\pi}+\sqrt{s}\right)-\log\left(\sqrt{\pi}\right)\right)$ correspondingly. 

Moreover, for an advanced civilization that exists on a theoretical Dyson's sphere \cite{dyson}, the surface area around the star is $4\pi {R}^2$, where $R$ is the spatial radius from s. As the civilization or the mega-structure metropolis grows on the shell encircling the star, its radius $R$ increases and the cells 
expand on the two-dimensional spherical surface. To reiterate, the number of cells, all of unit area, grow proportionally to  the surface area. With simple manipulation, we obtain 
\begin{equation} \label{eq9xDyson}
n\geq s\left(1 + \frac{4\pi}{\sqrt{s}}\right)\left(\log\left(4{\pi}+\sqrt{s}\right)-\log\left(4{\pi}\right)\right).
\end{equation}
In all three metropolis shapes that span two-dimensional surfaces, we can deduce that the metropolis surface area will continue to grow unless the following happens 
\begin{equation} \label{eq9xgeneral}
n\leq \frac{s}{2}\log\left(s\right).
\end{equation}
We have a cell growth equilibrium if the inequality above is satisfied. The area along the metropolis periphery then is smaller than the average empty area within. An equivalent interpretation of this equilibrium is that the metropolis perimeter becomes less than or equal to the maximal average perimeter of unoccupied cells inside the city. For instance, with square metropolises, the former is $4 \sqrt{s}$ whereas the latter can be no more than $4s \exp\left(-\frac{n}{s}\right)$ on average. The relationship between these perimeters can be shown to also reduce to $n\leq \frac{s}{2}\log\left(s\right)$. \emph{The area growth towards the exterior cannot happen since there is too much unoccupied area within the interior.}

\subsection{Population Growth Equilibrium}
Next we consider population growth given a constant metropolis area between successive epochs. As in Conway's \emph{cellular automaton} \cite{ConwayLife}, we postulate that the population will only increase if there are enough cells which are \emph{critically-populated}; cells that are neither under-populated nor over-populated. Between successive epochs, given a fixed $s$, the population $n$ may increase to any larger value as long as there will be some minimum number of cells on average that will be critically-populated. 

We define the average number of under-populated cells as ${\mathcal{E}}[L_0]$ and over-populated cells ${\mathcal{E}}[L_j]:j\geq 2$. Thus, the average number of critically-populated cells are ones with a single population unit (i.e. ${\mathcal{E}}[L_1]$). While other thresholds may also be calibrated for critical population, this serves as a useful starting point. The average number of critically-populated cells is then 
\begin{equation}
\label{eq2x2}
\begin{split}
A_1&=s\frac{(n/s)^{1}}{1!}\exp(-\frac{n}{s})\\
&=n\exp\left(-\frac{n}{s}\right),
\end{split}
\end{equation}
which is the mean number of cells with a single population unit. Increases in the metropolis population will only occur if this is sufficiently large. The maximum possible increase in $n$ between successive epochs can be generalized such that the average number of critically-populated cells is at least $\frac{s}{k}$. \emph{In other words, $\frac{1}{k}$ of the metropolis area on average must be neither under-populated nor over-populated to support further population growth.} When this is not satisfied we have ${\mathcal{E}}[L_1]\leq \frac{s}{k}$, where $1 \leq k\leq s$,
\begin{equation} 
\label{eq10}
\begin{split}
A_1&\leq \frac{s}{k}\\
\exp\left(-\frac{n}{s}\right)& \leq \frac{s}{k}\\
n&\leq s\left(\log\left(\frac{n}{s}\right) +\log{k}\right).
\end{split}
\end{equation}
Population growth equilibrium will be reached if the above inequality is satisfied, where the average number of critically-populated cells falls below $\frac{1}{k}$ of the metropolis area. At minimum, for population growth, there must always be at least one cell on average that is critically-populated, where the alternative to \eqref{eq10} simplifies to 
\begin{equation}\label{popgrowtheq}
n\geq s\log(n).
\end{equation}

\subsection{Growth Cycles}
Cell growth and population growth equilibria may occur in successive epochs. For instance, starting from some $(n^{(1)},s^{(1)})$ in an epoch, there may be cell growth to $s^{(2)}$ until equilibrium is reached (i.e. $n^{(1)}\leq \frac{s^{(2)}}{2}\log\left(s^{(2)}\right)$). In the next epoch, given the updated state $(n^{(1)},s^{(2)})$, the population grows until we achieve $n^{(2)}\leq s^{(2)}\log(n^{(2)})$. Thus, the cycles may alternate between cell growth and population growth over the epochs. 

We illustrate the concept of metropolis area (cell) growth given a high enough population density. In Fig \ref{fig:furt1}, we depict two population distribution ensembles of a metropolis with $s=256$ cells. With $n=867$ population units which randomly move between the cells, the population density  is $\lambda = \frac{n}{s}\approx 3.4$ with an average of $9$ unoccupied as per \eqref{eq9x}. Also note that as per \eqref{eq10}, the pair $(867,256)$ represents a population growth equilibrium for $k=10$ or or $10 \%$ area threshold.

Given a constant $n$, the area scales proportionally to the perimeter until we have $s=400$ as shown in Fig. \ref{fig:furt2}(a). We then have a $\lambda \approx 2.2$ population units with an average of $46$ unoccupied cells. At this point, any further cell growth by one cell along the city's perimeter would add $40$ new cells,  which is lower than the average number of unoccupied cells (i.e. 46). The cell growth reaches equilibrium at this point. 

In reality, both cell and population growth may happen simultaneously. The inequalities \eqref{eq9x} and \eqref{eq10} therefore represent Pareto-optimal boundaries between cell growth and population growth equilibria where only one or the other happen. In Fig. \ref{fig:furt2}b, we plot the relationship between endogenous growth of population and a square-shaped metropolis area by setting \eqref{eq9x} and \eqref{eq10} to equations. We obtain pairs of $(n,s)$ that satisfy the equations $n= s(1+\frac{4}{\sqrt{s}}+\frac{4}{s})\left(\log\left(4+\sqrt{s}\right)-\log(4)\right)$ and $n= s\left(\log\left(\frac{n}{s}\right) +\log{k}\right)$ respectively through a numerical method. Note that any coordinate $(n,s)$ that lies above the former equation represents a point which cannot experience cell growth as the population density is too low. Conversely, if $(n,s)$ lies below the latter equation cannot have an population increase as there are too few a critically-populated cells. 

Fig. \ref{fig:furt2}(b) is a log-linear plot that shows the narrow region (shaded white) between the two equations where both population and cells can increase ($\frac{k}{s}=0.1$ or $10 \%$ of the metropolis area). The transition between two equilibria can is shown for the coordinates denoting the pairs of $(n,s)$ for two distinct epochs. The line labeled \emph{sparsity threshold} represents the relationship in \eqref{eq8x}, where the average number of occupied cells equals the average number of unoccupied cells. 
 
\begin{figure}[t]
     \subfloat[\label{subfig-1:dummy}]{%
       \includegraphics[width=0.45\textwidth]{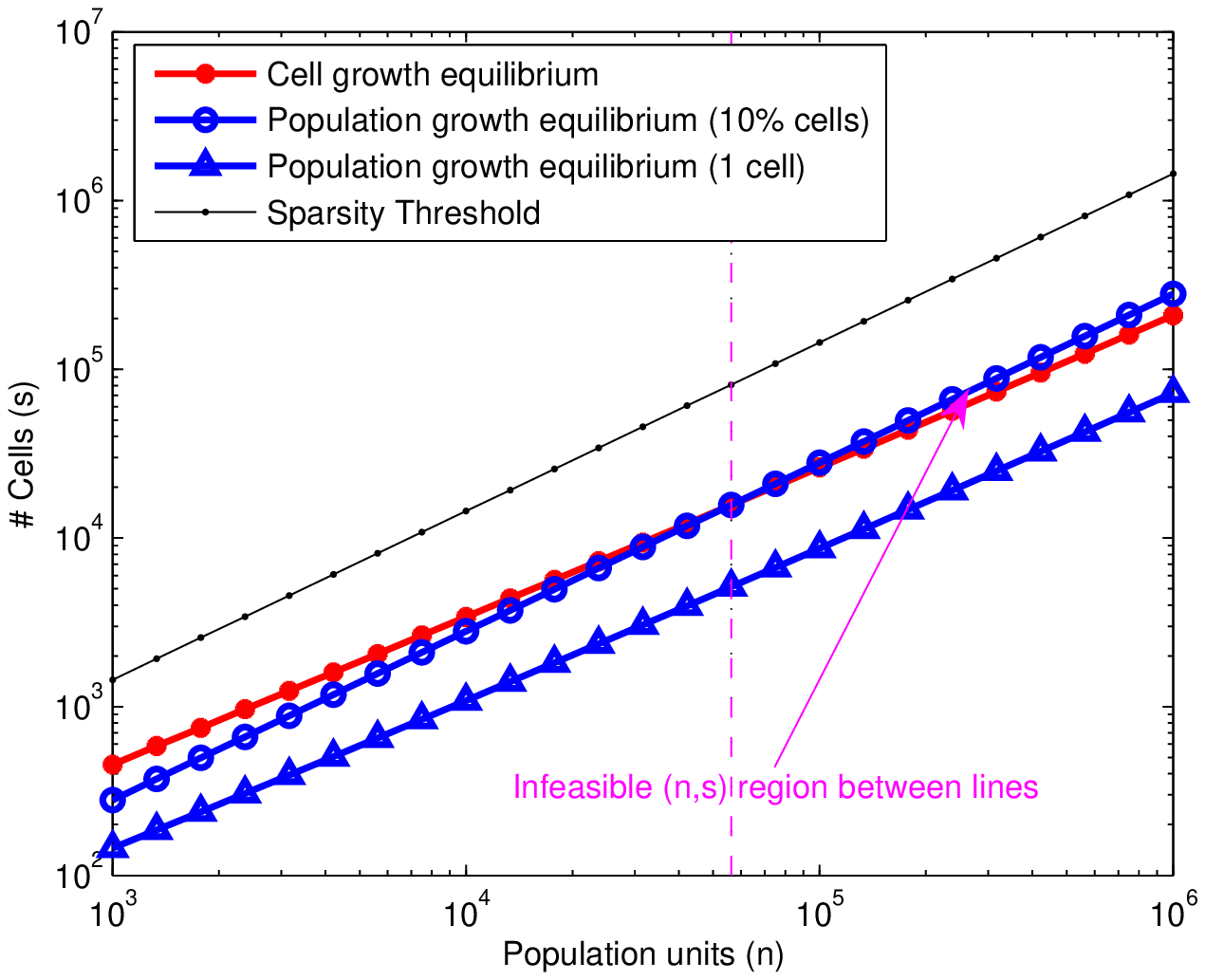}
     }
     \hfill
     \subfloat[\label{subfig-2:dummy}]{%
       \includegraphics[width=0.45\textwidth]{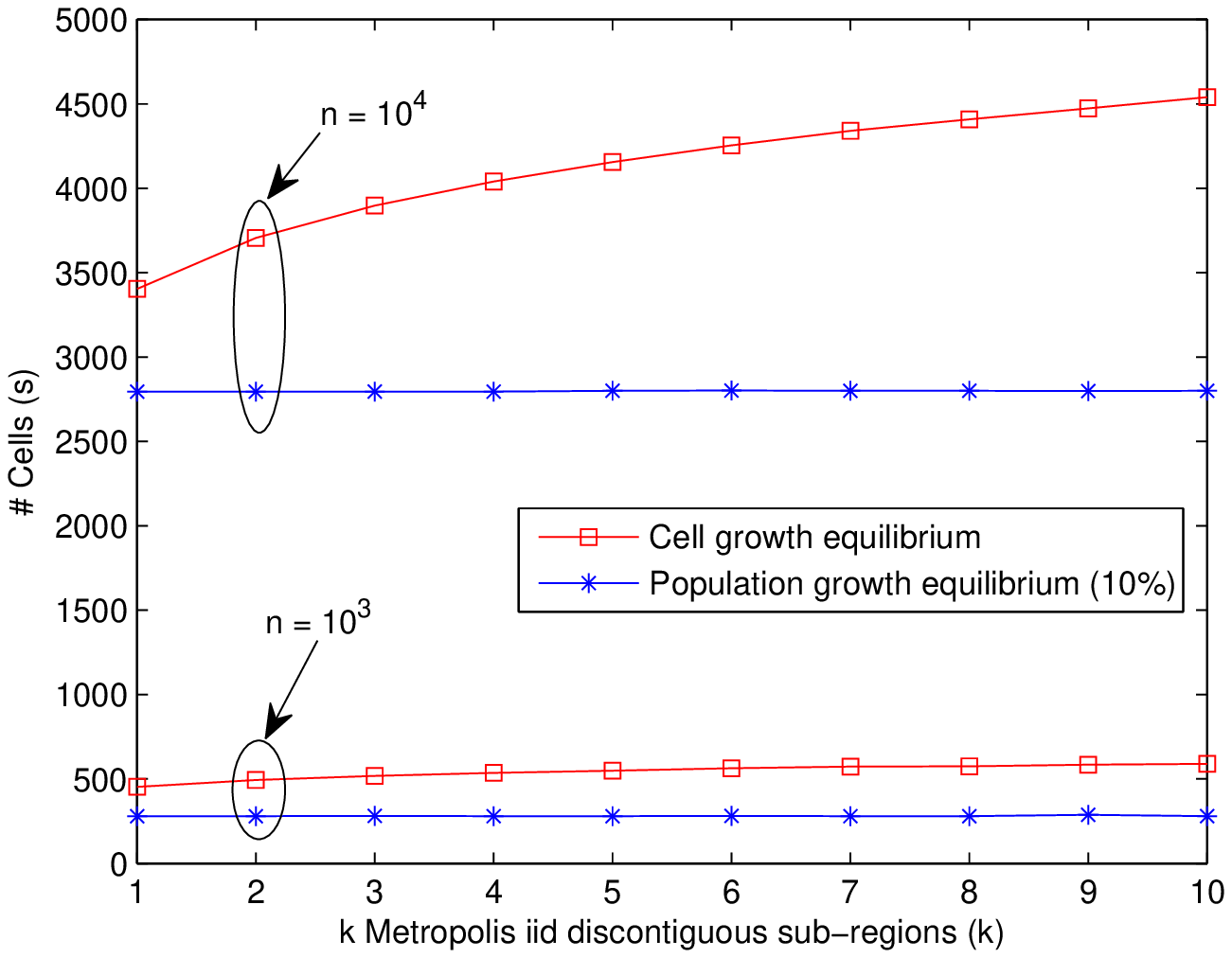}
     }
     \caption{(a) A contrast between equilibria points. (b) With discontiguous sub-regions, there is additional cell growth due to their larger aggregate perimeters as compared to a contiguous region. The population growth equilibrium is invariant to this factor.}
     \label{fig:furt2c}
   \end{figure}

The average number of critically-populated cells, which are deemed neither over-populated nor under-populated, determine how much increase in $n$ is possible. In the log-log plots in Fig. \ref{fig:furt2c}(a), for population growth equilibrium, the threshold for critically-populated cells is is set to $10 \%$ of $s$ as well as to just a single cell. When $n$ is large enough, the population growth Pareto-optimal equation becomes the largest for $10 \%$. Conversely, if a single cell is needed to be critically-populated on average, the equation has a lower gradient and does not intersect the cell growth equilibrium line. 

There is also another interesting implication for the coordinates in Fig. \ref{fig:furt2c}(a) for large populations (i.e. $n>10^{4.75}$) when the line for the population growth equilibrium at $10 \%$ is higher than that for the cell growth. Sandwiched between the two lines, these coordinates represent an infeasible set of pairs $(n,s)$. The coordinates below the former line denote over-crowded population and coordinates above the latter line indicate too much unoccupied area within the metropolis for cell growth. Hence, for these values neither kind of growth is possible. Conversely, if the population growth threshold is based on a fixed number of critically-populated cells then this restriction does not apply.

\subsection{Discontiguous Sub-region Growth}
In Fig. \ref{fig:furt2c}b, we plot the relationship between $n$ and $s$ assuming an equality constraint (i.e. exact equilibria) set in \eqref{eq9x} and \eqref{eq10} for a square metropolis area assuming that the city comprises $\kappa$ isolated and discontiguous sub-regions. In this case, the variable $\kappa: 1 \leq \kappa\leq s$ represents a sub-scaling factor denoting the number of smaller discontiguous sub-regions, each with population $\frac{n}{\kappa}$ and with $\frac{s}{\kappa}$ cells self-similarly organized as square sub-regions. {Each of these sub-regions will exhibit the same growth cycles between epochs as described earlier.} We presume that none of these $\kappa$ sub-regions will physically overlap as their areas grow and shall always remain discontiguous. Modifying \eqref{eq9x} for the case where (i.e. sub-scaled by $\kappa$), we have 
\begin{equation} 
\begin{split}\label{eq9aISO}
\frac{n}{\kappa}&\geq k\left(\frac{s}{\kappa}k\log(1-1/k)\right)\\
n&\geq \left(s(1+\frac{4}{\sqrt{s/\kappa}}+\frac{4}{s/\kappa})\left(\log\left(4+\sqrt{s/\kappa}\right)-\log(4)\right)\right)
\end{split}
\end{equation} 
Conversely, with $\kappa$ isolated sub-regions for the constraint on population growth, we observe that modifying \eqref{eq10} yields
\begin{equation} 
\label{eq10ISO}
\begin{split}
\frac{n}{\kappa}\left(\exp\left(-\frac{n/\kappa}{s/\kappa}\right)\right)&\geq \left(\frac{s/\kappa}{k}\right)\\
n&\geq s\left(\log\left(\frac{n}{s}\right) +\log{k}\right).
\end{split}
\end{equation}
Fig. \ref{fig:furt2c}b shows that with increasingly more sub-regions(or $\kappa$), the pairs $(n,s)$ under cell growth equilibria have a higher aggregate $s$, where the population per sub-region are at $\frac{n}{k}$. However, the converse is not true. The population growth equilibrium does not change. This means that when the sub-regions are isolated and discontiguous, there can be no extra population growth. However, with cell growth for each discontiguous sub-region, the aggregate area gained by all sub-regions is more than if all these sub-regions were a part of a single contiguous metropolis. To reinforce, this is due to the fact that the aggregate area of the perimeters of the $k$ isolated sub-regions is more than that of a single contiguous region.


\section{Energy Statistics}
Next we derive the statistics of the consumed energy of the metropolises defined by the pair $(n,s)$ in any given epoch. Recall that $\lambda=n/s$ is the population density. A closed-form expression of a probability mass function (pmf) of consumed energy is difficult to derive, owing to the number of variables involved. Hence, we simplify the problem by considering the mean and variance of the consumed energy of a single cell and scaling these metrics by the total number of cells to get a bound on the variance of the metropolis. Dividing \eqref{eq2} by $s$ gives us the average fraction of the time a given cell is occupied by $j$ population units. Using this approach, the mean consumed energy of the metropolis is
\begin{equation} 
\label{metromean}
{\bf{E}}\left[\mathcal{E}\right] =s \left(\sum_{j=0}^t \left(\frac{E_{\max}-E_{\min}}{t}j\right)\frac{\lambda^j\exp(-\lambda)}{j!} + E_{\max}\left(\sum_{j=0}^t 1- \frac{\lambda^j\exp(-\lambda)}{j!}\right)\right),
\end{equation}
where ${\bf{E}}\left[L_o\right]=s\exp\left(-\lambda\right)$.
The corresponding variance in the consumed energy of the metropolis is
\begin{equation} 
\label{eq3b}
\mathrm{Var}\left({{\mathcal{E}}}\right) = {\bf{E}}\left[{\mathcal{E}}^2\right] - \left({\bf{E}}\left[\mathcal{E}\right]\right)^2. 
\end{equation}
This can be expanded as
\begin{equation} 
\label{eq4}
\mathrm{Var}\left({{\mathcal{E}}}\right) =
{\bf{E}}\left[\left(\sum^{t-1}_{j=0}L_j\left(\frac{E_{\max}-E_{\min}}{t}.j + E_{\min}\right)+\sum^{s}_{j=t}L_jE_{\max}\right)^2\right] - \left({\bf{E}}\left[\mathcal{E}\right]\right)^2.
\end{equation}
The variance of the metropolis consumed energy is constrained by the variance of the consumed energy of any given cell multiplied by the total number of cells $s$ as follows
\begin{equation} 
\label{metrovar}
\begin{split}
\mathrm{Var}\left({{\mathcal{E}}}\right) \leq 
s \times {\Bigg[} \underbrace{ \sum_{j=0}^t \left(\frac{E_{\max}-E_{\min}}{t}j\right)^2\frac{\lambda^j\exp(-\lambda)}{j!} + E^2_{\max}\left(\sum_{j=1}^t 1- \frac{\lambda^j\exp(-\lambda)}{j!}\right) }_{\text{second moment cell energy}}\\
-\left(\underbrace{ \sum_{j=0}^t \left(\frac{E_{\max}-E_{\min}}{t}j\right)\frac{\lambda^j\exp(-\lambda)}{j!} + E_{\max}\left(\sum_{j=0}^t 1- \frac{\lambda^j\exp(-\lambda)}{j!}\right) }_{\text{first moment cell energy}}\right)^2{\Bigg]}.
\end{split}
\end{equation}
For $t=1$, we obtain a simple expression for the mean in \eqref{metromean} as
\begin{equation}\label{eq5}
{\bf{E}}\left[{\mathcal{E}}\right]=s\left(\left(E_{\min}-E_{\max}\right)\exp\left(-\lambda\right)+E_{\max}\right)
\end{equation}
and the bound on the variance in \eqref{metrovar} as
\begin{equation} 
\label{eq6}
\begin{split}
&\mathrm{Var}\left({{\mathcal{E}}}\right) \leq s{\Big[}\left(E^2_{\min}-E^2_{\max}\right)\exp\left(-\lambda\right) + E^2_{\max}{\Big]}- \\
&{\Big[}{\Big(} \left(E_{\min}-E_{\max}\right)^2\exp\left(-2\lambda\right)  -2E_{\max}\left(E_{\min}-E_{\max}\right)\exp(-\lambda) +E^2_{\max}    {\Big)} {\Big]}\\
&\leq s\left[ \left(E^2_{\min}+E^2_{\max}\right)\exp\left(-\lambda\right) - \left(E^2_{\min}+E^2_{\max}\right)\exp\left(-2\lambda\right) -2E_{\min}E_{\max}\exp\left(-\lambda\right)   \right]\\
&\mathrm{Var}\left({{\mathcal{E}}}\right) \leq s\left( \exp\left(-\lambda\right) - \exp\left(-2\lambda\right) \right)\left(E_{\min}-E_{\max}\right)^2
\end{split}
\end{equation}
Above represent elegant relationships connecting the variance of consumed energy to the number of cells in the metropolis, its population density, and the minimum and maximum energy per cell. At $t=1$, the mean consumed energy and the bound on the variance in \eqref{eq5} and \eqref{eq6} is combined as follows
\begin{equation} 
\label{eqVarViaMean}
\begin{split}
&\exp\left(-\lambda\right)= \frac{\frac{{\bf{E}}\left[{\mathcal{E}}\right]}{s} - E_{\max}}{E_{\min}-E_{\max}} \\
&\mathrm{Var}\left({{\mathcal{E}}}\right) \leq s \left[\frac{{\bf{E}}\left[{\mathcal{E}}\right]/s - E_{\max}}{E_{\min}-E_{\max}}\right]\left[1-\frac{{\bf{E}}\left[{\mathcal{E}}\right]/s - E_{\max}}{E_{\min}-E_{\max}}\right]\left(E_{\min}-E_{\max}\right)^2.
\end{split}
\end{equation}
This leads to a straight-forward quadratic relationship between the mean consumed energy and the variance of the metropolis as 
\begin{equation} 
\label{eqVarViaMeanBound}
\mathrm{Var}\left({{\mathcal{E}}}\right) \leq \frac{\left({\bf{E}}\left[{\mathcal{E}}\right]- sE_{\max}\right)\left(sE_{\min} - {\bf{E}}\left[{\mathcal{E}}\right]\right)}{s}.
\end{equation}

\begin{figure}[t]
     \subfloat[\label{subfig-1:dummy}]{%
       \includegraphics[width=0.45\textwidth]{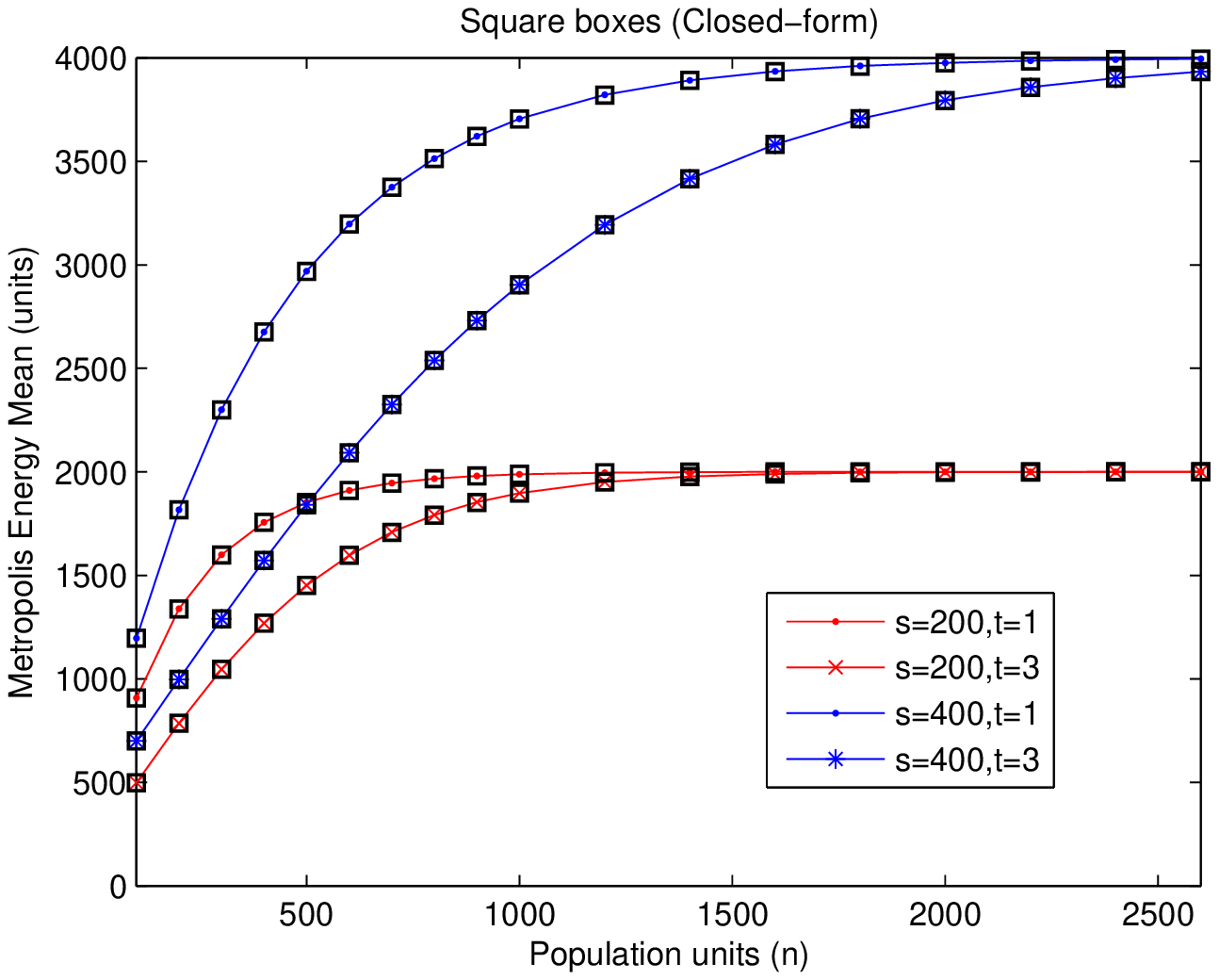}
     }
     \hfill
     \subfloat[\label{subfig-2:dummy}]{%
       \includegraphics[width=0.45\textwidth]{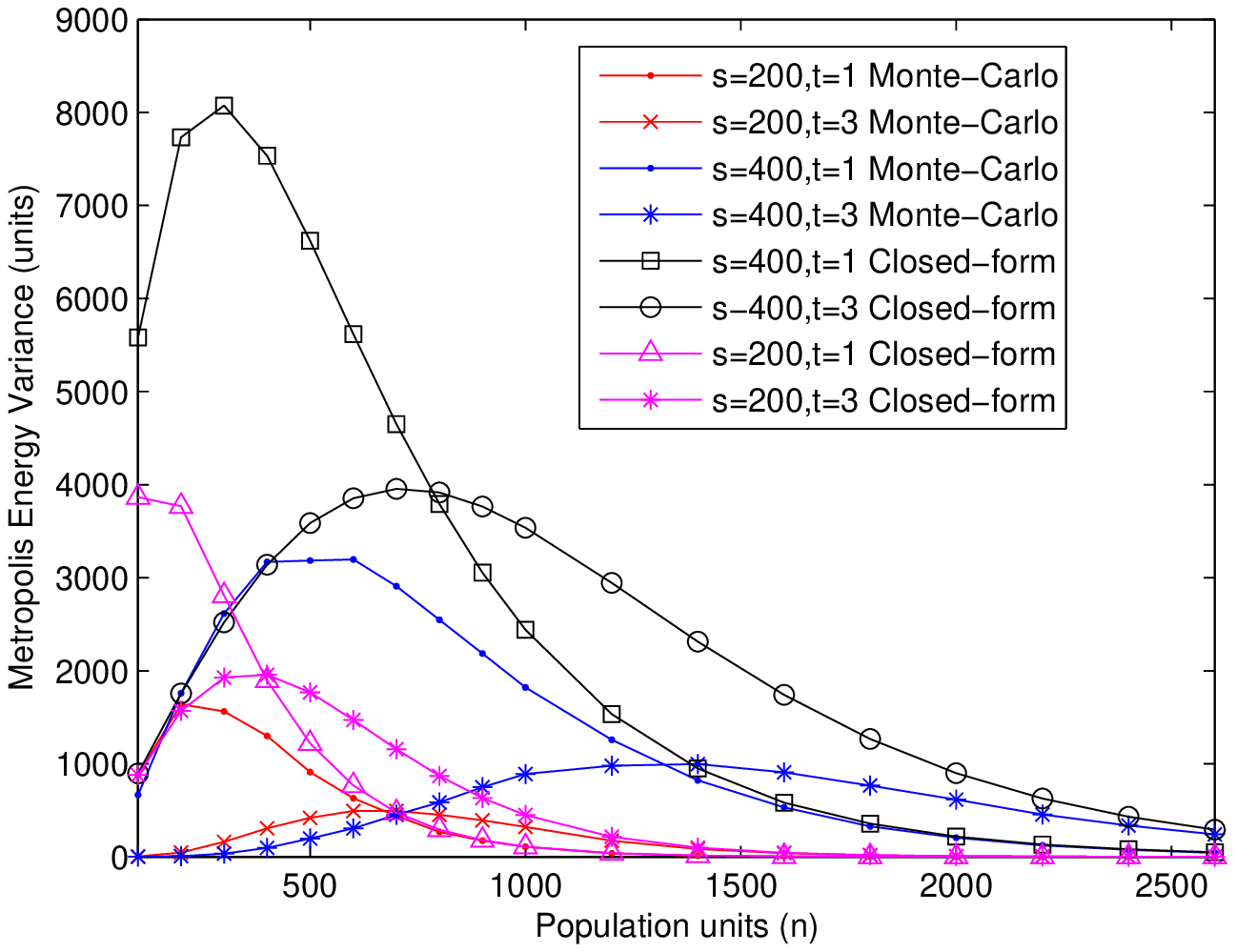}
     }
     \caption{Given $s$, the mean metropolis energy converges to the maximum for larger values of $n$ in (a), whereas in (b), the corresponding variance in the consumed energy first peaks and then gradually declines to zero.}
     \label{fig:furt3}
   \end{figure}

In subsequent results, we use the Monte-Carlo method to generate the statistics for consumed energy. By generating a large number of random samples with $n$ population units uniformly placed in the $s$ cells, the method provides the desired energy statistics. The energy parameters set in our Monte-Carlo trials are $E_{\min}=1$ and $E_{\max}=10$ units of energy. These values are chosen to reflect that the maximum energy any resource cell consumes may be an order of magnitude higher than the minimum. 

In Fig. \ref{fig:furt3}, we plot the mean and the variance of the consumed energy using Monte-Carlo method for various pairs of $(n,s)$. For a fixed metropolis area, as denoted by a constant $s$, we evaluate these statistics for for various population sizes of $n$ ranging from $100$ units to $2600$ units. We also have two values of the saturation threshold $t$ for illustration. In Fig. \ref{fig:furt3}(a), we observe that for small values of $n$, the consumed energy scales linearly before asymptotically converging to the peak threshold of $sE_{\max}$. Thus, with increasing $n$ we have mean consumed energy as ${\bf{E}}[{\mathcal{E}}]=4000$ for $s=400$ and ${\bf{E}}[{\mathcal{E}}]=2000$ for $s=200$. 
The scaling of variance in the consumed energy however exhibits a concave shape which diminishes to zero for large $n$ for both values of $s$ in Fig. \ref{fig:furt3}(b).  A higher saturation threshold at $t=3$ naturally tends to shift the curves, where the mean and variance of the consumed energy in both in Fig. \ref{fig:furt3}(a) and in Fig. \ref{fig:furt3}(b) reach their maximum at higher values of $n$.

In Fig \ref{fig:furt3}(a), the mean derived in \eqref{metromean}, represented by square coordinates, perfectly matches the mean consumed energy obtained through Monte-Carlo method. In Fig. \ref{fig:furt3}(b), the inequality in \eqref{metrovar} illustrates the bound on the variance of consumed energy.

Hence, we observe non-monotonicity and concavity of the variance of the consumed energy for fixed $s$ over the range of values of $n$. In other words, if the metropolis area is constant, smaller populations have less randomness in the consumed energy. With much larger populations, there will likely be multiple population units occupying a majority of cells. In this latter case, the consumed energy saturates for most cells, leading to a decline in the variance. 

\subsection{Consumed Energy and Economic size}
The size of the metropolis economy can also be indicated by its overall consumed energy \cite{Karanfil_ecoElec,rand_enisil}. Thus, the MDP (metropolis domestic product) can be hypothesized to be linearly proportional to the latter. Let ${\bf{E}}\left[{\mathcal{E}}^{(1)}\right]$ and ${\bf{E}}\left[{\mathcal{E}}^{(2)}\right]$ denote the mean consumed energy in two different epochs. Allowing $M_1$ and $M_2$ to then denote the magnitudes of their MDPs respectively, we may state that
\begin{equation}\label{MDPEN}
\frac{M_2}{M_1} = \frac{{\bf{E}}\left[{\mathcal{E}}^{(2)}\right]}{{\bf{E}}\left[{\mathcal{E}}^{(1)}\right]}.
\end{equation}
Suppose that the population density $\lambda=\frac{n^{(1)}}{s^{(1)}}=\frac{n^{(2)}}{s^{(2)}}$, is constant between the two epochs and that there is growth in the metropolis area by a factor of $k: k \geq 1$ (i.e. $s$ to $ks$). We surmise that
${\bf{E}}\left[L_j\right] =
ks\frac{\lambda^{j}}{j!}\exp\left(-\lambda\right)$. In this case, by modifying equation \eqref{eq3}, we have
\begin{equation} 
{\mathcal{E}}^{(2)}=\left(\sum^{t-1}_{j=0}kL_j\left(\frac{E_{\max}-E_{\min}}{t}.j + E_{\min}\right)+\sum^{s^{(2)}}_{j=t}kL_jE_{\max}\right)= k{\mathcal{E}}^{(1)}.
\end{equation}
Thus, with constant population density, as the number of resource cells increase from $s$ to $ks$, \eqref{MDPEN} simplifies to
\begin{equation}\label{MDPEN2}
\frac{M_2}{M_1} = k \frac{{\bf{E}}\left[{\mathcal{E}}^{(1)}\right]}{{\bf{E}}\left[{\mathcal{E}}^{(1)}\right]}=k.
\end{equation}
In other words, given a constant population density, the average consumed energy consumed scales linearly with $s$. \emph{Thus, the growth in MDP is then simply proportional to the overall increase in the mean consumed energy or equivalently the gain in the metropolis area.}

\section{System Entropy}
The energy consumed by each of the $s$ cells can be in $t+1$ energy states. Let the pmf of the consumed energy of the metropolis area be defined by $p_{\mathcal{E}}(z)$, where $z$ denotes the $z^{th}$ state from among the $s^{t+1}$ possible discrete or quantized consumed energy levels for the entire metropolis. The pmf can also be obtained through the Monte-Carlo method. 

The Shannon entropy \cite{coverthomas} of the consumed energy is a function of variables $n$, $t$ and $s$ such that
\begin{equation} 
\label{eq11}
H(n,s)=-\sum_{z=1}^{s^{t+1}}p_{\mathcal{E}}(z).\log\left(p_{\mathcal{E}}(z)\right)
\end{equation}
Mutual information between any two sub-regions represents how similar are the two zones in terms of consumed energy. The work in \cite{Met7} provides an interesting exposition using entropy to observe the dependencies of various sub-components of a large system. 

In this context, we consider the metropolis comprising $k$ sub-regions of equal area. If these sub-regions are discontiguous and isolated, with $\frac{n}{k}$ population units (iid), the consumed energy entropy of each sub-region is notated as $H(\frac{n}{k},\frac{s}{k})$. The entropy of the metropolis is then simply the sum of the entropies of all sub-regions
\begin{equation} 
\label{eq13}
H(n,s)= \sum_{i}^k H\left(\frac{n}{k},\frac{s}{k}\right).
\end{equation}
However, when the $k$ sub-regions are part of one contiguous region, where all $n$ population units can move to any cell, then the entropy of each sub-region is denoted as $H(n,s_1)$. The consumed energy of a sub-region is correlated to the consumed energy of the remaining sub-regions owing to greater population mobility. The mutual information between any one of the sub-regions and the other remaining $k-1$ sub-regions in the contiguous region is defined as
\begin{equation} 
\label{eq12}
I(s_1,s_{k-1})= H(n,s_1)+H(n,s_{k-1})-H(n,s).
\end{equation}
Note that in above $H(n,s_k)$ is the entropy of sub-region $1$ (identical for all $k$ sub-regions) and $H(n,s_{k-1})$ is the entropy of the set of the $k-1$ remaining sub-regions. Note that in the discontiguous case, the mutual information between any two sub-regions is always zero. 


\begin{figure}[t]
     \subfloat[\label{subfig-1:dummy}]{%
       \includegraphics[width=0.45\textwidth]{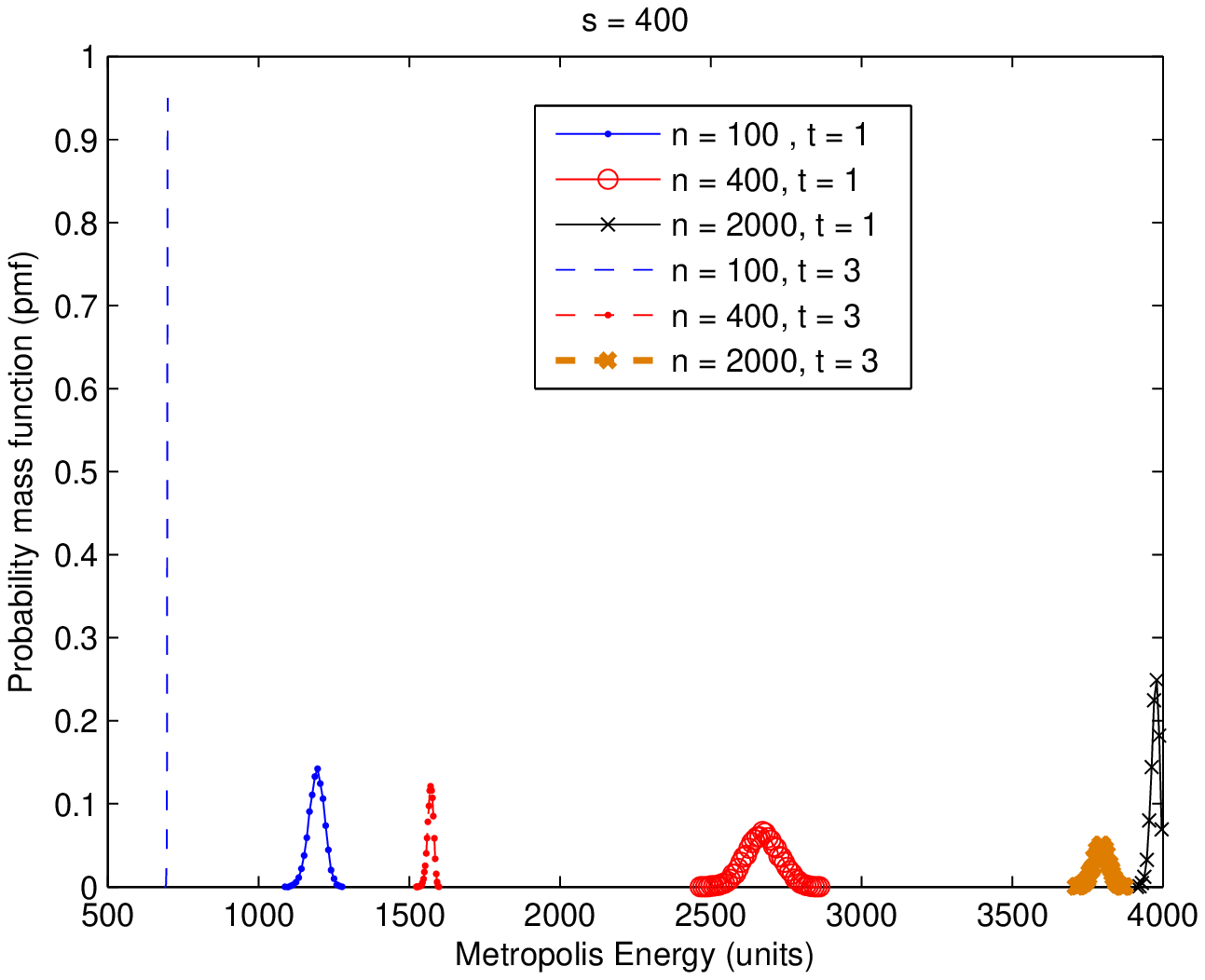}
     }
     \hfill
     \subfloat[\label{subfig-2:dummy}]{%
       \includegraphics[width=0.45\textwidth]{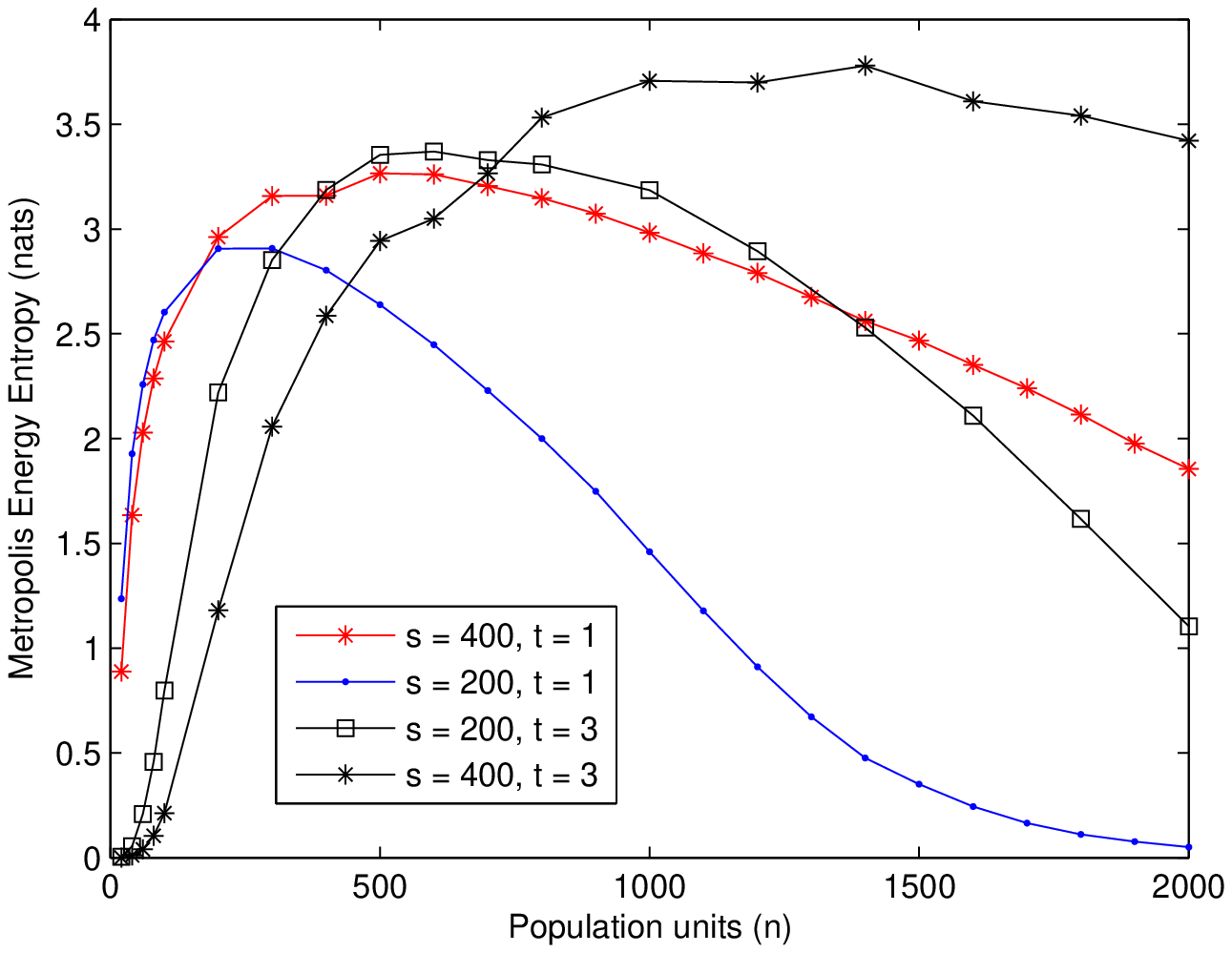}
     }
     \caption{(a) The pmf of the consumed energy with various population thresholds. (b) The entropy of the metropolis shows a similar trend as the variance in Fig. \ref{fig:furt3}b.}
     \label{fig:furt6}
   \end{figure}


\begin{figure}[t]
     \subfloat[\label{subfig-1:dummy}]{%
       \includegraphics[width=0.45\textwidth]{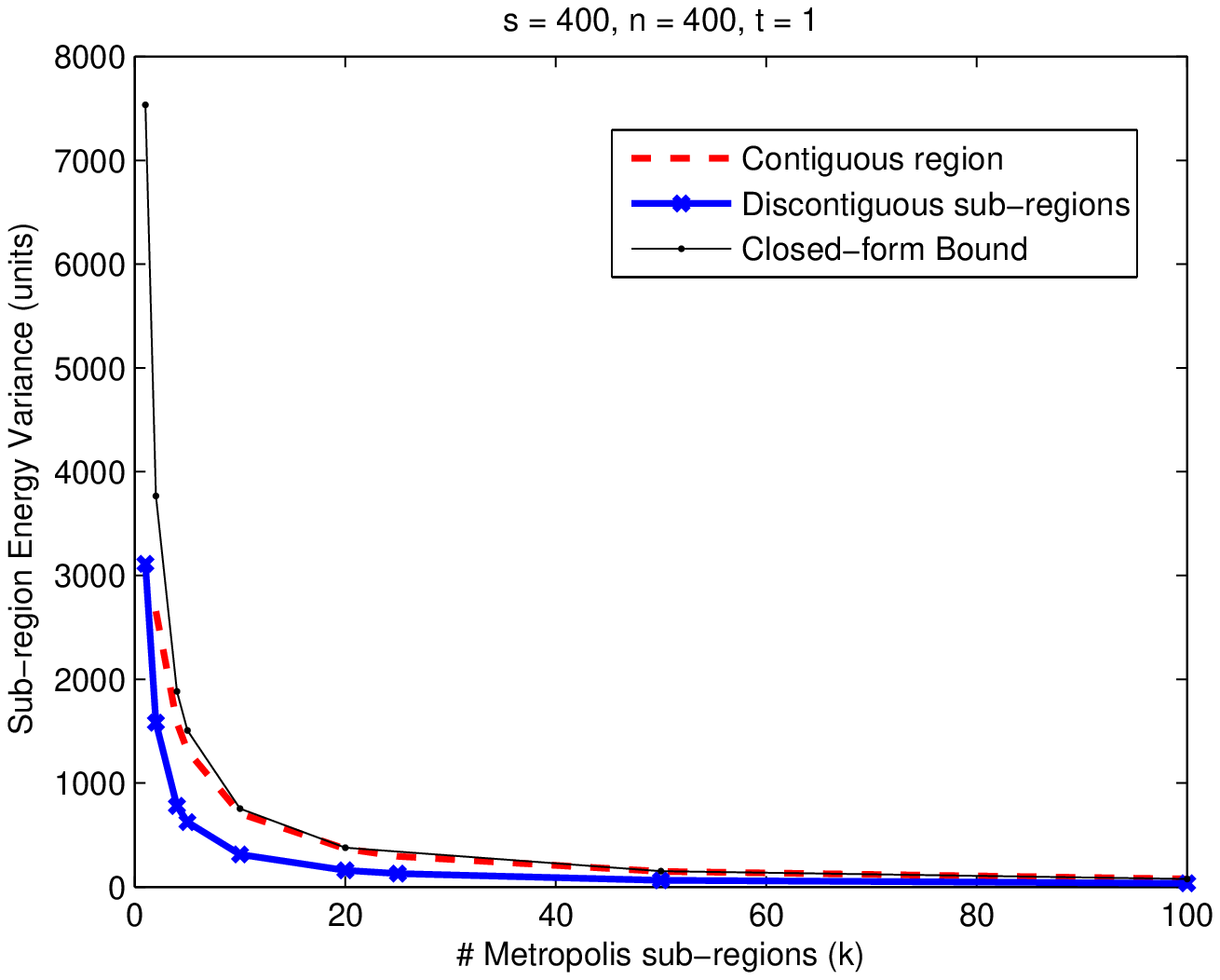}
     }
     \hfill
     \subfloat[\label{subfig-2:dummy}]{%
       \includegraphics[width=0.45\textwidth]{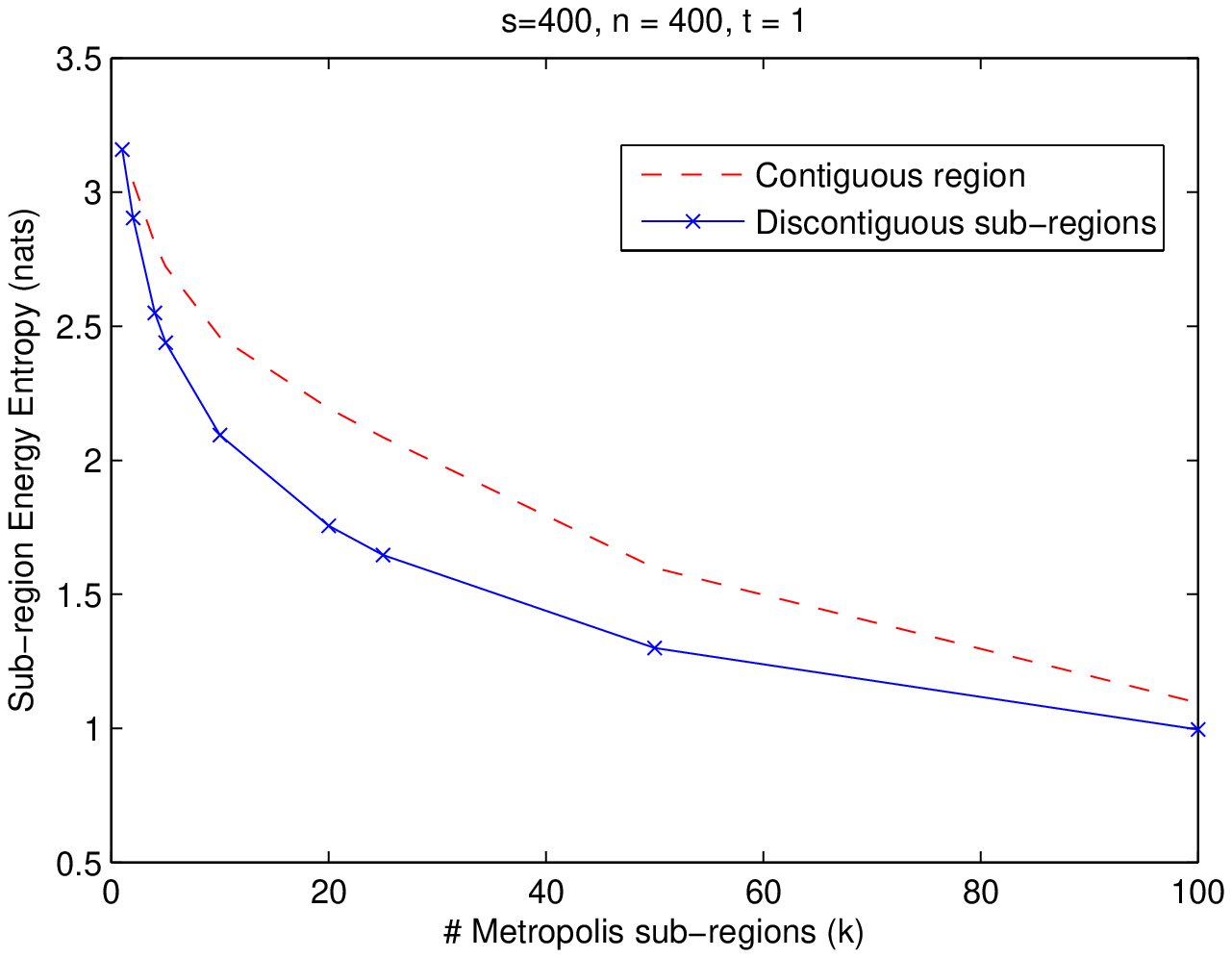}
     }
     \caption{The (a) variance and (b) entropy of any single sub-region of the metropolis. With discontiguous sub-regions, these metrics are lower than when there is a contiguous metropolis.}
     \label{fig:furt7}
   \end{figure}
   
\begin{figure}[!t]
\begin{center}
        \centering\includegraphics[width=0.47\textwidth]{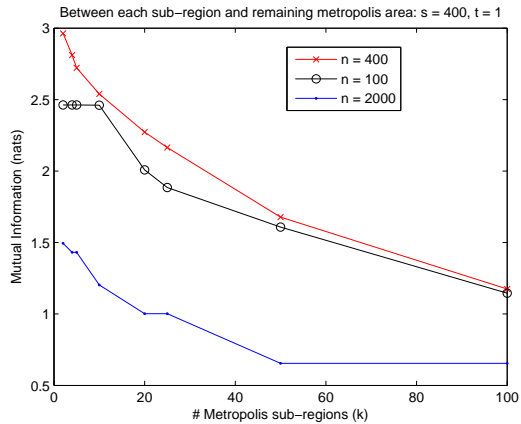}
   \caption{The mutual information between any one sub-region and the remaining $k-1$ sub-regions declines with $k$. When $k=2$, we essentially have maximum similarity of consumed energy pattern of the two large metropolis sub-regions.}
\label{fig:furt8}
\end{center}
\end{figure}

In Fig. \ref{fig:furt6}(a), we plot the pmf of the consumed energy for three values of $n$. We can observe that the spread of the pmf, which essentially corresponds to the variance in consumed energy, is (1) small for $n=100$, (2) increases for the intermediate case $n=400$ and (3) eventually diminishes again for $n=2000$. In Fig. \ref{fig:furt6}(b), we plot the entropy \eqref{eq11} of the consumed energy over the range of $n$ for two thresholds of both $s$ and $t$. We observe that the maximum entropy occurs at the same values of $n$ where the variance in the consumed energy became maximum in Fig. \ref{fig:furt3}(b).

In Fig. \ref{fig:furt7}, we plot the variance and entropy of a single sub-region both in the contiguous and discontiguous settings. Note that $k=1$ corresponds to the default case with one sub-region. We can see that with increasing $k$, these metrics decline quickly both situations. As the population units are free to migrate across all region in the contiguous setting, we observe a higher variance and entropy as compared to when the population is isolated into discontiguous pockets.  \emph{It implies that variance and entropy of the metropolis consumed energy indicate the spatial mobility or spatial freedom of the population}. \emph{Arbitrary mobility patterns of the population would thus only produce a lower variance or entropy as compared when the population can move between all cells.} 

In Fig. \ref{fig:furt8} plots the the mutual information between any single sub-region and the remaining $k-1$ sub-regions for different values of $k$. Essentially, we illustrate how representative is a given sub-region of the remaining metropolis area. When $k=2$, the metropolis has two sub-regions, where naturally the mutual information between them is also high. When $k=100$, the mutual information diminishes significantly for any given $n$ as any sub-region is one-hundredth of the metropolis area and provides a poor approximation of the consumed energy pattern for the remaining $99$ sub-regions. Our results also reinforce the results in \cite{Met7} that the marginal utility of information declines with scale. 

\section{Conclusion}
Complexity can be found in social networks, biological, engineering or economic systems. We have proposed a new approach to study the scaling laws, entropy and consumed energy of a metropolis that is a complex system comprising all these components. Using an abstract model of a metropolis in terms of two fundamental variables consisting of population and area (or cells more broadly), we have shown that when the metropolis area is fixed, increasing population initially increments consumed energy's variance or entropy up to a certain threshold and which then declines afterwards. Moreover, for the case when population and cells can mutually affect, we have derived equations that define the Pareto-optimal boundaries where either population increase or metropolis area expansion happen. We have also shown that variance or entropy of the consumed energy of a metropolis can represent tha spatial freedom of a mobile population. Future work will consider applying these principles by calibrating them to any actual data available for various real-world metropolises. Our model may serve as a blueprint to examine broad classes of complex systems.

\begin{figure}[!t]
\begin{center}
        \centering\includegraphics[width=0.47\textwidth]{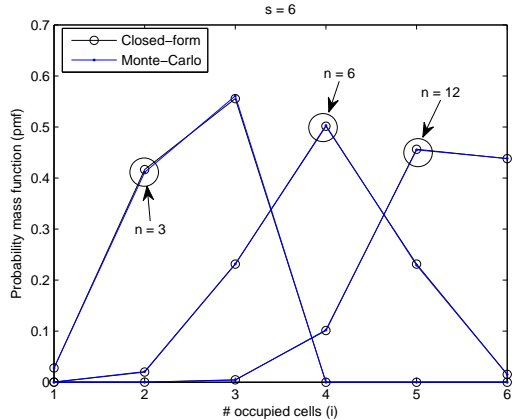}
\caption{ The Monte-Carlo approach validates the expression in \eqref{eqPMF}, which is the pmf of the number of occupied cells and where the latter is computationally-intensive to obtain.}
\label{fig:Occupypmf}
\end{center}
\end{figure}

\section*{Acknowledgement}
The author would like to appreciate the review and comments by Dr. Abdul Naseer Malmi Kakkada of University of Texas-Austin.

\appendix
Given the pair $(n,s)$, we derive the pmf of the number of cells occupied by at least one population unit through an iterative algorithm approach. As an illustration, consider a small system with $n=4$ and $i=2$ occupied cells. We iteratively put all the $4$ population units in the cells randomly one by one such as only $i=2$ cells are occupied. In each iteration, let an event success $S$ be defined as when a population unit is put in a cell that is currently unoccupied and failure $F$ as when the population unit is put in a cell that already has a population unit. The very first population unit will always have success as no cell is initially occupied. With additional population units randomly placed in the cells, the probability of failure will increase as more and more of the cells are likely occupied. Specifically, for $i=2$, we will have the following iteration combinations $SSFF$, $SFFS$ and $SFSF$. Note that
the corresponding probability of $i$ occupied cells for each combination is $\frac{s}{s}.\frac{s-1}{s}.\frac{2}{s}.\frac{2}{s}$, $\frac{s}{s}.\frac{1}{s}.\frac{1}{s}\frac{s-1}{s}$ and $\frac{s}{s}.\frac{1}{s}.\frac{s-1}{s}.\frac{2}{s}$, respectively. We can infer that the probability of $i$ successes is thus $\prod_{j=1}^{i-1}\left(1-\frac{j}{s}\right)$. However, the probability of failures depends on the prior set of occupied cells as population units are placed in them in succession. The first $i$ population units should randomly go in any $i$ unoccupied cells. The remaining $n-i$ population units then must fall in these $i$ cells in successive iterations.

Since there are $i$ successes, let $a_k: k \in \{1,2,\cdots,i\}$ be the number of consecutive failures between successive successes such that $SF^{a_1}SF^{a_2}\cdots SF^{a_i}$ and where $\sum_{k=1}^i a_k = n-i$. We can deduce that
the pmf that $i$ cells are occupied by at least one population unit is 
\begin{equation} 
\label{eqPMF}
P_i=\prod_{j=1}^{i-1}\left(1-\frac{j}{s}\right)\sum_{a_1=0}^{n-i}\sum_{a_2=0}^{n-i}\cdots \sum_{a_{i}=0}^{n-i }\prod_{k=1}^{i}\left(\frac{k}{s}\right)^{a_m}:\sum_{m=1}^{i} a_m = n-i.
\end{equation}
There are $i$ summation terms in \eqref{eqPMF} with the constraint that the total number of failures equals $\sum_{k=1}^i a_k = n-i$. The computational complexity of obtaining the pmf of the number of occupied cells above is prohibitive. As the pmf of the consumed energy depends on the distribution of occupied cells, a Monte-Carlo method is a quicker and efficient approach to determining $p_{\mathcal{E}}(z)$. In Fig. \ref{fig:Occupypmf}, we plot the pmf for various pairs of $(n,s)$ and validate the expression through a Monte-Carlo approach.

\bibliography{MetroBibliography}

\end{document}